%% file: main.tex
\documentclass[10pt,letterpaper,nofonttune]{IEEEtran}

\newcommand{\bs}{\boldsymbol}

\newcommand{\bigO}[1]{\mathcal{O}(#1)}

\usepackage[numbers,sort&compress]{natbib}
\usepackage{url}  
\usepackage{amsthm}
\usepackage{amssymb}
\usepackage{amsfonts}
\usepackage{amsmath}
\usepackage{eso-pic}
\usepackage{graphicx}
\usepackage{tikz}
\usepackage{pgf}
\usepackage{lscape}
\usepackage{fancyvrb}
\usepackage{verbatim}
\usepackage{xfrac}
\usepackage{algpseudocode,algorithm,algorithmicx}
\usepackage{multirow}
\usepackage{proof}
\usepackage{epstopdf}
\usepackage{subfigure}
\usetikzlibrary{arrows,automata}
\usetikzlibrary{shapes}

\newcommand{\bss}{\mathcal{B}}
\newcommand{\mvnos}{\mathcal{M}}

\newcommand{\rbs}{\mathcal{R}}
\newcommand{\nrb}{N_{RB}}
\newcommand{\slicing}{\mathbf{L}}
\newcommand{\intf}{\mathbf{Y}}

\newcommand{\reffig}[1]{Fig.~\ref{#1}}   

\theoremstyle{plain}
\newtheorem{theorem}{Theorem}

\newtheorem*{corollary*}{Corollary}

\theoremstyle{definition}
\newtheorem{definition}{Definition}
\newtheorem*{definition*}{Definition}

\theoremstyle{remark}

\newtheorem*{remark*}{Remark}

\title{\textit{The Slice Is Served:} Enforcing Radio Access Network Slicing in Virtualized 5G Systems}
\author{\IEEEauthorblockN{
Salvatore D'Oro\IEEEauthorrefmark{1},
Francesco Restuccia\IEEEauthorrefmark{1},
Alessandro Talamonti\IEEEauthorrefmark{2},
and Tommaso Melodia\IEEEauthorrefmark{1}}\\
\IEEEauthorblockA{\IEEEauthorrefmark{1}Department of Electrical and Computer Engineering, Northeastern University, Boston, USA, \\ Email: \{s.doro, f.restuccia, t.melodia\}@northeastern.edu.}\\
\IEEEauthorblockA{\IEEEauthorrefmark{2}Politecnico di Milano, Milan, Italy, Email: alessandro.talamonti@mail.polimi.it.}
\thanks{This  work  was supported in part by ONR under Grant 0014-16-1-2213 and Grant N00014-17-1-2046 and in part NSF under Grant CNS-1618727.}
\thanks{This paper has been accepted for publication in IEEE INFOCOM 2019. This is a preprint version of the accepted paper. Copyright (c) 2019 IEEE. Personal use of this material is permitted. However, permission to use this material for any other purposes must be obtained from the IEEE by sending a request to pubs-permissions@ieee.org.}
}

\begin{document}

\maketitle
\pagenumbering{gobble} 

\input{abstract.tex}
\input{introduction.tex}
\input{related_work}
\input{system_model}
\input{problem_statement}

\input{our_solution}
\input{numerical_analysis}
\input{experimental_analysis}
\input{conclusions}

\footnotesize
\bibliographystyle{IEEEtran}
\bibliography{acmart} 

\input{appendix_shorter}

\end{document}

%% file: abstract.tex
\begin{abstract}
The notions of softwarization and virtualization of the radio access network (RAN) of next-generation (5G) wireless systems are ushering in a vision where applications and services are physically decoupled from devices and network infrastructure. This crucial aspect will ultimately enable the dynamic deployment of heterogeneous services by different network operators over the same physical infrastructure. RAN slicing is a form of 5G virtualization that allows network infrastructure owners to dynamically ``slice" and ``serve" their network resources (\textit{i.e.}, spectrum, power, antennas, among others) to different mobile virtual network operators (MVNOs), according to their current needs. Once the slicing policy (\textit{i.e.}, the percentage of resources assigned to each MVNO) has been computed, a major challenge is how to allocate spectrum resources to MVNOs in such a way that (i) the slicing policy defined by the network owner is enforced; and (ii) the interference among different MVNOs is minimized. In this article, we mathematically formalize the RAN slicing enforcement problem (RSEP) and demonstrate its NP-hardness. For this reason, we design three approximation algorithms that render the solution scalable as the RSEP increases in size. We extensively evaluate their performance through simulations and experiments on a testbed made up of 8 software-defined radio peripherals. Experimental results reveal that not only do our algorithms enforce the slicing policies, but can also double the total network throughput when intra-MVNO power control policies are used in conjunction.
\end{abstract}

%% file: introduction.tex
\section{Introduction}\label{sec:intro}

Recent studies indicate that the demand for faster, lower-latency wireless cellular connection is growing exponentially each year. By 2023, Ericsson has predicted that around 3.5 billion cellular IoT connections will be active, and more than 1 billion 5G subscriptions will be activated worldwide \cite{EricssonMobility2018}. It is now clear that existing one-size-fits-all resource allocation policies will not be able to sustain the sheer need for dynamic, effective and efficient radio access strategies, where network operators will need to make the very best use of the extremely limited spectrum bands available for commercial usage  \cite{SpectrumCrunch}.

To address the above issues, radio access network (RAN) slicing has been recently welcomed as a promising approach by the academic and industrial communities alike \cite{nakao2017end,kokku2013cellslice,foukas2017orion,rost2017network}. RAN slicing completely overturns the traditional model of single ownership of all network resources, and realizes a vision where physical network infrastructure is shared among multiple mobile virtual network operators (MVNOs), each one in charge of a separate ``slice'' of the network, which can be assigned/revoked by the network owner according to the current slicing policy. Although companies such as Amazon or Microsoft apply similar concepts in the context of cloud computing, \emph{RAN slicing is an intrinsically different problem, since spectrum is a scarce resource for which over-provisioning is not possible}. 

Once the RAN slices have been defined for each MVNO, a core problem is how to divide and allocate the spectrum resources, also called resource blocks (RBs), according to what prescribed by the slicing policy. In other words, if (for example) an MVNO has been assigned a slice of 15\% of the spectrum resources, such MVNO should receive approximately 15\% of the available RBs. Thus, the design and evaluation of \textit{RAN slicing enforcement} algorithms is necessary to implement in practice the network owner (NO)'s slicing policy. Moreover, to be effective, RAN slicing enforcement algorithms must facilitate interference-mitigating strategies such as inter-base-station power control (IBSPC) \cite{hossain2014evolution,d2018learning}, MIMO \cite{jungnickel2014role}, and coordinated multi-point (CoMP) \cite{irmer2011coordinated,jungnickel2014role,lee2012coordinated} schemes such as Joint Transmission (JT) \cite{7848911,nam2014advanced}. However, since such technologies require tight cooperation and coordination among different base stations (BSs), \textit{it becomes imperative to design effective and efficient slicing enforcement algorithms to guarantee that the same (or similar in time/frequency) RBs are assigned to the same MVNOs when BSs are close enough to interfere among themselves.}

\begin{figure}[!h]
    \centering
    \includegraphics[scale=0.3]{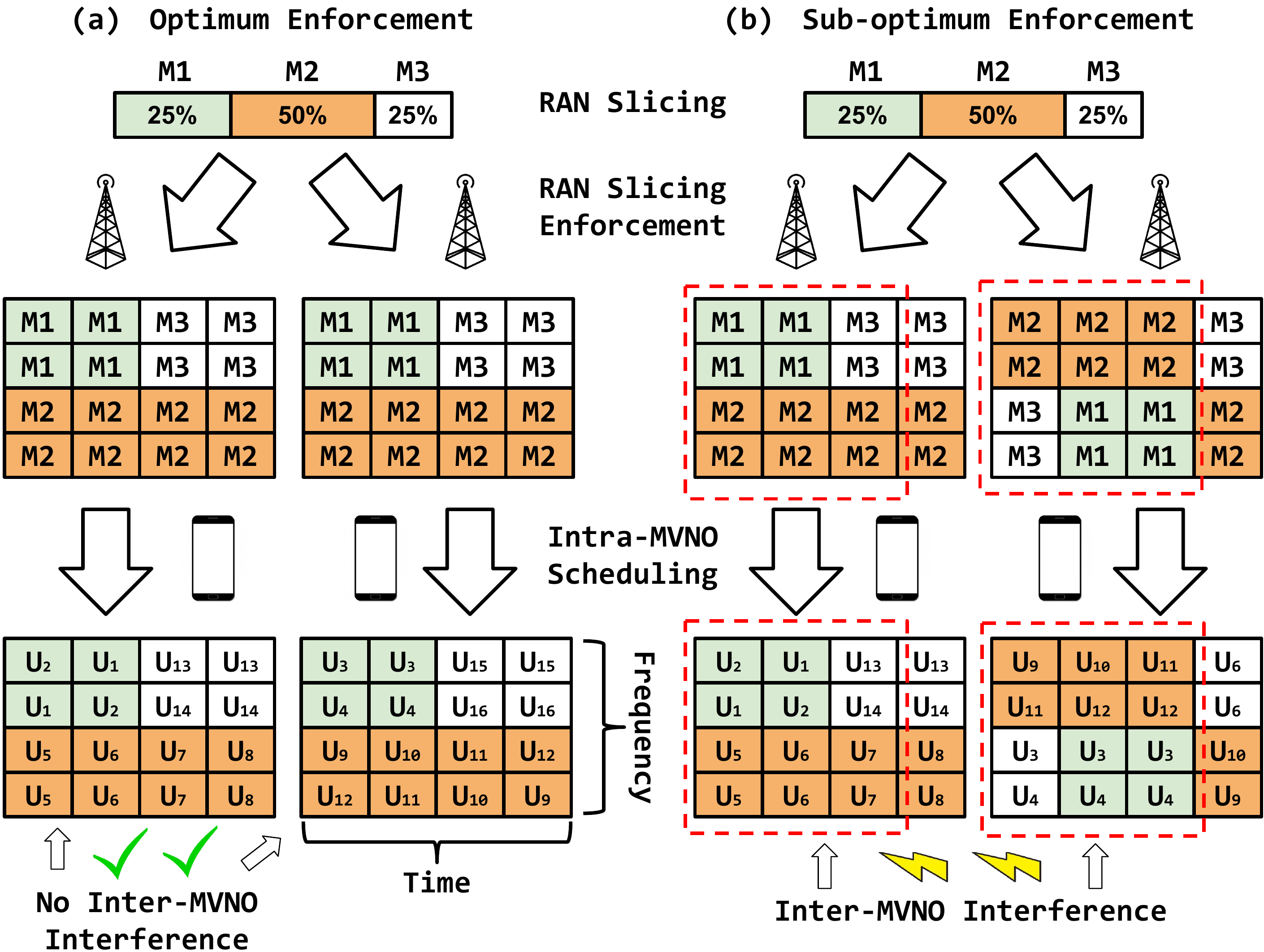}
    \caption{Optimum and Sub-optimum RAN Slicing Enforcement.}
    \vspace{-0.2cm}
    \label{fig:example_slicing}
 \end{figure}
 
To illustrate the above point, we consider the cellular network scenario depicted in Figure \ref{fig:example_slicing}. Here, the network owner administers two BSs (assumed to be close enough to interfere with each other) and 16 RBs (\textit{i.e.}, 4 frequency units during 4 time units). We consider the case where three MVNOs, namely M1, M2 and M3, have been assigned the following slice: $\mathrm{M1 = 25\%, M2 = 50\%, M3 = 25\%}$, on both the BSs. Figure \ref{fig:example_slicing}(a) shows an optimum slicing enforcement, represented as two \textit{RB allocation matrices} (RBAM), where inter-MVNO interference is absent (\textit{i.e.}, MVNOs control the same RBs at the two BSs). In this case, MVNOs have maximum flexibility and can easily mitigate interference between their mobile users (MUs) residing in the two BSs by using IBSPC. Conversely, the right side of Figure \ref{fig:example_slicing} shows sub-optimum RBAMs which cause inter-MVNO interference during 12 RBs.

\begin{figure}[h!]
\centering
   \begin{minipage}[b]{0.25\columnwidth}
    \centering
    \includegraphics[width=\columnwidth]{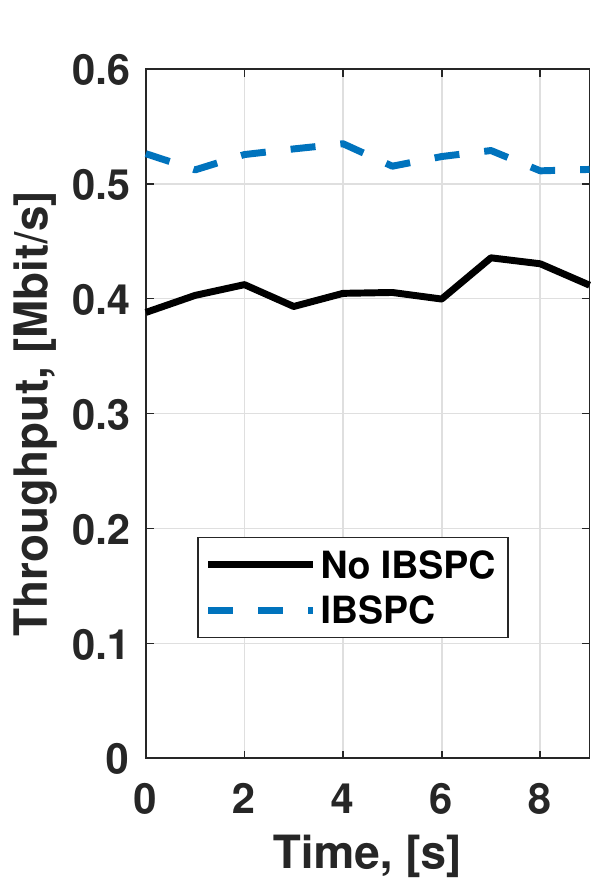}
    \caption{\label{fig:example_ibspc} Experimental Results, IBSPC vs No IBSPC.}
  \end{minipage}
  \hspace{0.01\textwidth}
  \begin{minipage}[b]{0.7\columnwidth}
    \includegraphics[width=\columnwidth]{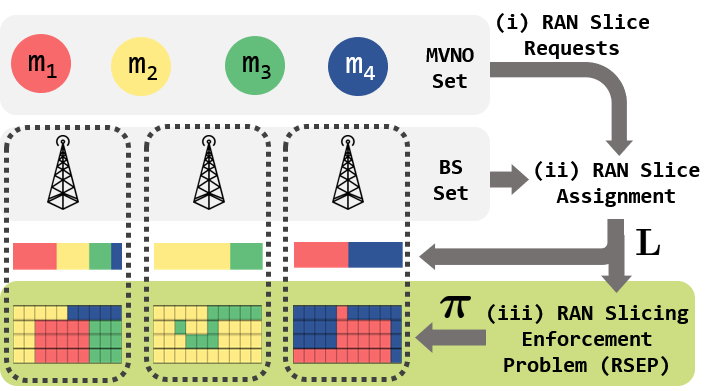}
    \caption{\label{fig:architecture} The RAN slicing architecture.}
  \end{minipage}
\end{figure}
To further investigate the issue of severe inter-MVNO interference, we ran a series of experiments on the testbed to be discussed in detail in Section \ref{sec:experimental}. In such experiments, the RBs are assigned randomly to MVNOs according to the percentage defined in the slicing policy. 
We performed channel estimation and then applied IBSPC to keep inter-cell interference below a fixed threshold.
Figure \ref{fig:example_ibspc}, which depicts the total network throughput as a function of time, indicates that by using IBSPC, the network can increase its throughput by about $25\%$.\textit{ In Section \ref{sec:experimental}, we show that an optimal slicing policy in conjunction with IBSPC can double the network throughput with respect to a random slicing enforcement algorithm.}

The issue of RAN resource allocation
has attracted large research interest over the last few years \cite{afolabi2018network,kaloxylos2018survey,7926919}. However, the problem of physical-level allocation of spectrum resources to MVNOs still remains underexplored, as discussed in details in Section \ref{sec:related} (Related Work). This is not without a reason; the design of slicing enforcement algorithms presents the following unique challenges, which are substantially absent in traditional RAN resource allocation scenarios: 

\begin{enumerate} 
\item \textit{Feasibility}: RBs need to be assigned to each MVNO according to the slicing policy. In other words, if an MVNO has been assigned 30\% of spectrum resources by the network owner, then it should receive 30\% of the total RBs available; 
\item \textit{Orthogonality}: at the same time, orthogonality among each RAN slice must be ensured. That is, each RB should be allocated to only one MVNO to avoid interference and other performance-degrading factors \cite{chang2018radio,foukas2017orion,sallent2017radio};
\item \textit{Enabling of advanced 5G technologies}: the RB allocation should facilitate advanced signal processing and RF transmission technologies such as IBSPC, JC, CoMP and MIMO, that considerably improve network performance and have been envisioned to lead the way of 5G networks.
\end{enumerate}

The objective of this work is to design, analyze and experimentally evaluate RAN slicing enforcement algorithms that address the three critical challenges mentioned above. Specifically, this paper makes the following contributions:

\begin{itemize} 
\item We formulate the \textit{RAN slicing enforcement problem} (RSEP) in Section \ref{sec:problem}, and show that its optimization version named \textit{RSEP-QP} is NP-hard in Section \ref{sec:optimal}. To this end, we propose (i) an approximated version of RSEP-QP, named \textit{RSEP-EQ}, in Section \ref{sec:approximation}; and (ii) a heuristic algorithm named \textit{RSEP-MLF} in Section \ref{sec:greedy};
\item  We prove in Section \ref{sec:reduc_complex} that the  execution time of both \textit{RSEP-QP} and \textit{RSEP-EP} can be drastically reduced (under specific conditions) by solving an equivalent problem, named the \textit{aggregated RSEP};  
\item We evaluate the performance of the proposed algorithms through extensive simulations (Section \ref{sec:numerical}) and through experimental evaluation on a testbed composed of 8 USRP software-defined radio devices (Section \ref{sec:experimental}). Results show that our approximation algorithms provide throughput close to the the optimal solution, and that by using IBSPC, our algorithms double the network throughput with respect to a baseline algorithm where slices are allocated in a random fashion. 
\end{itemize}



%% file: related_work.tex
\section{Related Work}\label{sec:related}

The issue of dividing RAN resources among a set of MVNOs has received significant interest from the research community over the last years; for excellent surveys on the topic, the reader is referred to \cite{afolabi2018network,kaloxylos2018survey}. This problem (also referred to as \textit{RAN slicing problem} \cite{7926919,8253541,foukas2017orion}) has been addressed by designing solutions that use a variety of theoretical tools, ranging from optimization \cite{zhao2018information,caballero2017multi,softslice,8370043}, auctions \cite{8057046} and game theory \cite{8377187,narmanlioglu2017learning,8532127}. 

The key limitation of the work above is that it does not show how to actually deploy RAN slices on top of the underlying physical network. For this reason, the enforcement of RAN slicing policies has attracted interest from the research community. Prior work \cite{foukas2017orion,7962822,rost2017network,ferrus20185g,Ksentini-ieeecommmag2017} provides high-level orchestration platforms that virtualize the available resources to create pools of resources that are then shared and allocated among the MVNOs. Although this approach is  efficient from a control and orchestration point of view, it might be ineffective in those scenarios where \textit{fine-grained control of physical-layer resources is required}, for example, to enable IBSPC, CoMP and beamforming transmissions.

A few recent papers have focused on addressing the RAN slicing enforcement problem from a resource allocation perspective. Chang \emph{et al.} \cite{chang2018radio} propose a partitioning algorithm that allocates the available RBs to each requesting MVNO by simultaneously maximizing the percentage of satisfied MVNOs while allocating the minimum amount of RBs. Similarly, Han \emph{et al.} \cite{8382171} consider genetic algorithms to assign the available RBs to the MVNOs such that a long-term utility is maximized. However, \cite{chang2018radio,8382171} analyze the problem considering a network with a single BS, and thus cannot be applied in multi-cell networks where MVNOs request different amounts of resources on different BSs. The authors in \cite{sallent2017radio,mahindra2013radio} identify fine-grained RB management as a promising approach to guarantee orthogonality and reduce inter-MVNO interference, thus deploying highly-efficient 5G networks. However, \cite{sallent2017radio} does not provide any algorithm to enforce slicing policies to maximize network efficiency, while \cite{mahindra2013radio} does not consider interference among BSs when allocating RBs.

\textit{To summarize, our work separates itself from existing literature on RAN resource allocation in the following aspects.} We first provide a novel formulation of the RAN slicing enforcement problem (RSEP) that (i) satisfies MVNOs requests, (ii) enforces orthogonality by reducing inter-MVNO interference, and (iii) enables advanced communication techniques such as CoMP and coordinated beamforming by maximizing the number of simultaneous MVNO transmissions on different BSs. Then, we propose three algorithms that solve the RSEP with different optimality performance and computational complexity, and we show how they can be applied to virtualized LTE networks through simulations. Finally, through a real-world experimental testbed, we (i) demonstrate the feasibility and effectiveness of the proposed solutions; (ii) show that our model efficiently reduces inter-MVNO interference by (a) enforcing orthogonality among slices and (b) increasing network throughput by enabling intra-MVNO transmission strategies. 

%% file: system_model.tex
\vspace{-0.1cm}
\section{System Model and Problem Overview} \label{sec:system}

We consider the RAN shown in \reffig{fig:architecture}, consisting of a set $\bss=\{1,2,\dots,B\}$ of $B$ base stations (BSs), each associated with a \textit{coverage area} $\rho_b,~b\in\bss$. Two BSs $b$ and $b'$ are \textit{interfering} (or \textit{adjacent}) with each other if $\rho_b\cap \rho_b' \neq \emptyset$. We define $\intf = (y_{b,b'})_{b,b'\in\bss}$ as a symmetric adjacency matrix such that $y_{b,b}=0$ for all $b\in\bss$, $y_{b,b'}=1$ if BSs $b$ and $b'$ interfere with each other, and $y_{b,b'} = 0$ otherwise.

 
The RAN is administered by a NO, who periodically rents to a set $\mvnos=\{1,2,\dots,M\}$ of $M$ MVNOs virtual RAN slices built on top of the underlying physical network $\bss$. For the sake of generality, we assume RAN slices are valid for $T$ time slots. This way, slow-changing networks (\textit{e.g.}, cellular networks in rural areas during nighttime) can be modeled with large values of $T$, while small values of $T$ can be used to model fast-changing networks (\textit{e.g.}, urban areas during daytime).

Similar to LTE, spectrum resources are represented as RBs, where each RB represents the minimum spatio-temporal scheduling unit \cite{dahlman20134g}. Also, we consider an OFDMA channel access scheme (as in downlink LTE) where  RBs are organized as a \textit{resource grid}, and where $\nrb$ and $T$ represent the number of available subcarriers and temporal slots, respectively. Thus, the set of available resources at each BS is $\rbs$, with $|\rbs| = \nrb \cdot  T$.

We explain the interaction between MVNOs and the NO with the help of \reffig{fig:architecture}. First, (i) MVNO's RAN slice requests are collected by the NO.
Then, (ii) the NO generates a \textit{slicing profile} $\slicing = (L_{m,b})_{m\in\mvnos,b\in\bss}$ where $L_{m,b}$ represents the amount of resources that the NO should allocate to MVNO $m\in\mvnos$ on BS $b$ in the time span $0 \le t \le T$ (\textit{i.e.}, \textit{RAN Slice Assignment}); and (iii) computes a slicing enforcement policy $\bs{\pi}$ that allocates RBs among the MVNOs such that all MVNO's requests are satisfied (\textit{i.e.}, \textit{RAN Slice Enforcement}).

%


This paper addresses point (iii), since (ii) has been already extensively investigated  \cite{afolabi2018network,kaloxylos2018survey,7926919,8253541,foukas2017orion,zhao2018information,caballero2017multi,softslice,8057046,8377187,narmanlioglu2017learning}. Specifically, we show how the NO can compute an efficient slicing enforcement policy $\bs{\pi}$ that satisfies the three requirements described in Section \ref{sec:intro}, \textit{i.e.}, feasibility, orthogonality and enabling of advanced technologies. 

%% file: problem_statement.tex
\section{The RAN Slicing Enforcement Problem (RSEP)}\label{sec:problem}

For any given slicing profile $\slicing$ and BS $b$, we identify the subset $\mvnos_b \subseteq \mvnos$ of MVNOs that include BS $b$ in their RAN slice as $\mvnos_b = \{m\in\mvnos : L_{m,b}>0 \}$.

Let $x_{m,b,n,t} \in\{0,1\}$ be the \emph{RB allocation indicator} such that $x_{m,b,n,t}=1$ if RB $(n,t)\in\rbs$ is allocated to MVNO $m$, $x_{m,b,n,t}=0$ otherwise.
Also, let $\bs{\pi} = (\bs{\pi}_b)_{b\in\bss}$ be the \textit{slicing enforcement policy}, where $\bs{\pi}_b = (\bs{\pi}_{m,b})_{m\in\mvnos}$ and $\bs{\pi}_{m,b}$ represents the set of RBs on BS $b$ that are allocated to MVNO $m$.
In more detail, for any RB $(n,t)\in\rbs$, we have that $(n,t)\in\bs{\pi}_{m,b} \iff x_{m,b,n,t}=1$. 
Hence, the set $\Pi$ of all feasible slicing enforcement policies $\bs{\pi}$ can be defined as:
\begin{align} \label{eq:feasbility_set}
    \Pi  & = \{ \bs{\pi} = (\bs{\pi}_{m,b})_{m\in\mvnos,b\in\bss} : |\bs{\pi}_{m,b}| = L_{m,b} ~\wedge \nonumber \\ 
         & \bs{\pi}_{m,b} \cap \bs{\pi}_{m',b} = \emptyset ~\forall m\neq m', m,m' \in \mvnos, ~b\in\bss \}
\end{align}

To properly formulate the RSEP, we now introduce the concept of linked RBs.
\begin{definition}[Linked RBs]
A given RB $(n,t)$ on the resource grid is \textit{linked} to MVNO $m$ on two interfering BS $b$ and $b'$ if and only if $x_{m,b,n,t}=x_{m,b',n,t}=1$ and $y_{b,b'}=1$.
\end{definition}

Linked RBs indicate those RBs that have been assigned to the same MVNO on adjacent BSs.
Specifically, a linked RB allows the corresponding MVNO to simultaneously access a specific spectrum portion in the same time-slot from two different BSs.

The reason why this feature is particularly relevant to the RSEP is threefold: (i) MVNOs can use linked RBs to enable advanced transmission schemes (\textit{e.g.}, power control, beamforming, MIMO and CoMP transmissions) among nearby BSs; (ii) as shown in \reffig{fig:example_slicing}, linked RBs can be used to deploy fully-orthogonal RAN slices with no inter-MVNO interference; and (iii) linked RBs do not generate inter-MVNO interference, thus avoiding any need for centralized coordination or distributed coordination among MVNOs.

It is clear that the maximization of the number of simultaneously linked RBs addresses the three issues identified in Section \ref{sec:intro}. Thus, we focus our attention on this approach. By leveraging the concept of linked RBs, we define $n_{b,b',m}$ as follows:
\begin{equation} \label{eq:number_interference}
    n_{b,b',m} = y_{b,b'}\cdot |\bs{\pi}_{m,b} \cap \bs{\pi}_{m,b'}|,
\end{equation}

where \eqref{eq:number_interference} represents the number of linked RBs among interfering BSs. It is also worth noting that the relationship $n_{b,b',m} = n_{b',b,m}$ always holds for all $b,b' \in \bss$ and $m\in\mvnos$.

For each MVNO $m\in\mvnos$, the total number $N_m$ of linked RBs on the corresponding RAN slice under policy $\slicing$ is
\begin{equation} \label{eq:number_1}
    N_m = \frac{1}{2} \sum_{b\in\bss} \sum_{b'\neq b} y_{b,b'}\cdot  n_{b,b',m},
\end{equation}
\noindent
where the $1/2$ factor is introduced to avoid double counting the same RBs. 

We formally define the RSEP as follows:
\begin{align} 
 \underset{\bs{\pi}\in\Pi}{\text{maximize}} & \hspace{1cm} \sum_{m\in\mvnos} N_m \label{prob:RSEP} \tag{RSEP}
\end{align}
\noindent

In a nutshell, the objective in RSEP is to compute a feasible slicing enforcement policy $\bs{\pi}$ that maximizes the total number of linked RBs while guaranteeing that the computed policy does not violate the feasibility constraint $\bs{\pi}\in\Pi$.

%% file: our_solution.tex
\vspace{-0.1cm}
\section{Addressing the RSEP Problem}\label{sec:solution}

\reffig{fig:example_slicing} shows that the formulation in Problem \ref{prob:RSEP} is particularly well-suited for RAN slicing problems. This is because it satisfies MVNOs requirements in terms of number of obtained RBs, helps orthogonality among slices through the reduction of inter-MVNO interference, and enables coordination-based communications such as CoMP, JT and beamforming.

To solve Problem \ref{prob:RSEP}, we need to compute a slicing enforcement policy by exploring the feasible set $\Pi$. Given the formulation in Problem \ref{prob:RSEP} does not in itself provide any intuitions on how a solution can be computed. For this reason, we (i) reformulate Problem \ref{prob:RSEP} by using the RB allocation indicators introduced in Section \ref{sec:problem}; (ii) show that the problem is NP-hard; and (iii) present a number of algorithms to compute both optimal and sub-optimal solutions to Problem \ref{prob:RSEP}.
\vspace{-0.1cm}
\subsection{Optimal Solution} \label{sec:optimal}

By using the definition of the RB allocation indicator $x_{m,b,n,t}$ and from \eqref{eq:feasbility_set}, we have that \eqref{eq:number_1} can be reformulated as
\begin{equation} \label{eq:nm_new}
    N_m = \frac{1}{2} \sum_{t=1}^T \sum_{n=1}^{\nrb} \sum_{b\in\bss} \sum_{b'\neq b, ~b'\in \bss} y_{b,b'} x_{m,b,n,t} x_{m,b',n,t}.
\end{equation}

Let us consider the matrices $\mathbf{B} = \intf \otimes \mathbf{I}_{\nrb \cdot T}$ and $\mathbf{Q} = \mathbf{I}_{M} \otimes \mathbf{B}$, where $\otimes$ stands for Kronecker product and $\mathbf{I}_k$ is the $k \times k$ identity matrix.
From \eqref{eq:nm_new}, it can be easily shown that $\sum_{m\in\mvnos} N_m = \frac{1}{2} \mathbf{x}^\top \mathbf{Q} \mathbf{x}$. Accordingly, Problem \ref{prob:RSEP} can be reformulated as
\begin{align}
 \underset{\bs{x}}{\text{maximize}} & \hspace{0.2cm} \frac{1}{2} \mathbf{x}^\top \mathbf{Q} \mathbf{x} \label{prob:RSEP-QP} \tag{RSEP-QP} \\
    \text{subject to} & \hspace{0.1cm} \sum_{t=1}^T \sum_{n=1}^{\nrb} x_{m,b,n,t} = L_{m,b}, \hspace{0.1cm} \forall b\in\bss, ~\forall m\!\in\!\mvnos   \label{constr:B1} \tag{C1} \\
                      & \hspace{0.1cm} \sum_{m\in\mvnos} x_{m,b,n,t} \leq 1, \hspace{0.6cm} \forall (n,t) \in \rbs, ~\forall b\in\bss    \label{constr:B2} \tag{C2} \\
                      & \hspace{0.1cm} x_{m,b,n,t}\!\!\in\!\! \{0,1\}, \forall (n,t)\! \in\! \rbs, \forall b\!\in\!\bss, \forall m\!\in\!\mvnos \label{constr:B3} \tag{C3}
\end{align}
\noindent
where $\mathbf{x}=(x_{m,b,n,t})_{m,b,n,t}$ can be represented as a $MB\nrb T \times 1 $ column array.

In Problem \ref{prob:RSEP-QP}, Constraint \eqref{constr:B1} ensures that all MVNOs receive the assigned number of RBs, while Constraint \eqref{constr:B2} guarantees that each RB is allocated to one MVNO only. Finally, Constraint \eqref{constr:B3} expresses the boolean nature of the RB allocation indicator. 
Problems \ref{prob:RSEP} and \ref{prob:RSEP-QP} are equivalent, as the latter is a reformulation of the former in terms of the RB allocation indicator. However, this new formulation shows that the RSEP can be modeled as a 0-1 (or binary) Quadratic Programming (QP) problem. We prove in Theorem \ref{th:np-hard} that Problem \ref{prob:RSEP-QP} is NP-Hard.

\begin{theorem} \label{th:np-hard}
Problem \ref{prob:RSEP-QP} is NP-hard.
\end{theorem}
\begin{IEEEproof}
It is sufficient to show that the matrix $\mathbf{Q}$ is \textit{indefinite}, \textit{i.e.}, it admits both positive and negative eigenvalues. Indeed, it is well-known \cite{pardalos1991quadratic,sahni1974computationally} that even real-valued non-binary QP problems are NP-hard when $\mathbf{Q}$ is indefinite.

From the definition of $\mathbf{B}$ and $\mathbf{Y}$, matrix $\mathbf{Q}$ has all zero entries in the main diagonal. Accordingly, $\mathbf{Q}$ is a zero-diagonal (or hollow) symmetric matrix. Let $\bs{\lambda}$ be the set of eigenvalues of $\mathbf{Q}$. Notice that $\sum_{\lambda_i\in\bs{\lambda}} \lambda_i= \mathrm{Tr}\{\mathbf{Q}\}$, and $\mathrm{Tr}\{\mathbf{Q}\}=0$ in our case. Thus, all the eigenvalues of $\mathbf{Q}$ must sum up to zero, meaning that either all eigenvalues are equal to zero, or $\mathbf{Q}$ has both positive and negative eigenvalues. Thanks to the symmetry of $\mathbf{Q}$, we can exclude the former case since it would imply that $\mathbf{Q}$ is the zero-matrix (\textit{i.e.}, there is no interference among BSs and $y_{b,b'}=0$ for all $b,b'\in\bss$). Therefore, $\mathbf{Q}$ must have both positive and negative eigenvalues, \textit{i.e.}, $\mathbf{Q}$ is indefinite. This proves the theorem.
\end{IEEEproof}


Since Problem \ref{prob:RSEP-QP} is NP-hard, in Section \ref{sec:approximation} we leverage  linear relaxation and  the concept of equivalence to design a reduced-complexity solution to Problem \ref{prob:RSEP-QP}, while in Section \ref{sec:greedy} we design a heuristic algorithm that can compute a sub-optimal solution to Problem \ref{prob:RSEP-QP} with polynomial complexity.
\vspace{-0.1cm}
\subsection{Approximated Solution} \label{sec:approximation}

Let $V = M\cdot B \cdot \nrb \cdot T$, and let us consider the following transformed problem 
\begin{align}
\underset{\bs{x}}{\text{maximize}} & \hspace{0.2cm} \frac{1}{2} \mathbf{x}^\top (\mathbf{Q} + 2\lambda \mathbf{I}_V) \mathbf{x} - \lambda \bs{e}^\top \mathbf{x} \label{prob:RSEP-EQ} \tag{RSEP-EQ} \\
    \text{subject to} & \hspace{0.2cm} \eqref{constr:B1},\eqref{constr:B2} \nonumber \\
                & \hspace{0.1cm}  0 \!\leq\! x_{m,b,n,t} \!\leq\! 1, \forall (n,t) \!\in\! \rbs, \!\forall b\!\in\!\bss, \!\forall m\!\in\!\mvnos \label{constr:E3} \tag{C4}
\end{align}
\noindent
where $\lambda\in\mathbb{R}$ is a real-valued parameter whose relevance to Problem \ref{prob:RSEP-EQ} will be explained in Theorem \ref{th:equivalence}, and $\bs{e}^\top=(1,1,\dots,1)$. The following theorem holds.

\begin{theorem} \label{th:equivalence}
There exists $\lambda\in\mathbb{R}$ such that Problem \ref{prob:RSEP-EQ} is equivalent to Problem \ref{prob:RSEP-QP}. Also, let $z^*$ be the largest (positive) eigenvalue of $\mathbf{Q}$. For any $\lambda\geq-z^*$, Problem \ref{prob:RSEP-EQ} is a quadratic convex problem over the unit hypercube.
\end{theorem}
\begin{IEEEproof}
Notice that $\mathbf{Q}$ contains only $0$-$1$ entries and $ x_{m,b,n,t} \leq 1$, which implies that $\mathbf{x}^\top \mathbf{Q} \mathbf{x}$ is always bounded and finite. Also,  $\mathbf{x}^\top \mathbf{Q} \mathbf{x}$ has continuous and bounded first-order derivatives over the unit hypercube, \textit{i.e.}, it is Lipschitz-continuous in any open set that contains the unit hypercube. From \cite[Th.~3.1]{Giannessi1999}, it must exist $\lambda_0\in\mathbb{R}$ such that $\forall \lambda\geq \lambda_0$ Problems \ref{prob:RSEP-EQ} and \ref{prob:RSEP-QP} are equivalent. Intuitively, the utility function in Problem \ref{prob:RSEP-EQ} introduces the term $\lambda \mathbf{x}^\top(\bs{e} - \mathbf{x})$ which generates a cost, or a penalty, proportional to $\lambda$ when constraint $x_{m,b,n,t}\in\{0,1\}$ is not satisfied. Accordingly, the binary constraint in Constraint \eqref{constr:B3} can be dropped and relaxed with the unit hypercube constraint $0\leq x_{m,b,n,t}\leq 1 $.
Recall that $\mathbf{Q}$ matrix admits both negative and positive eigenvalues. Accordingly, let $\mathbf{z}$ be the set of eigenvalues of $\mathbf{Q}$ and $z^*=\max\{z_1,z_2,\dots,z_{|\mathbf{z}|}\}$, we can choose $\lambda \geq z^* $ to show that the matrix $\mathbf{Q} + 2\lambda \mathbf{I}_V$ is positive semi-definite. Thus, Problem \ref{prob:RSEP-EQ} is convex, which proves the Theorem.
\end{IEEEproof}

\textbf{Remarks.} Theorem \ref{th:equivalence} shows that it is possible to relax the binary constraint of Problem \ref{prob:RSEP-QP} by replacing it with a penalty term. This produces an equivalent convex QP problem where binary variables $x_{m,b,n,t}$ are replaced with continuous ones through a linear relaxation. 
In general, local and global solutions of convex quadratic maximization problems (and the corresponding concave quadratic minimization problems) lie on the vertices of the feasibility set \cite{floudas1995quadratic}, thus making Problem \ref{prob:RSEP-EQ} easier to  solve when compared to Problem \ref{prob:RSEP-QP}. In some cases, Problem \ref{prob:RSEP-EQ} might still require exponential time with respect to the number of vertices. Approaches that restrict the search space to the vertices of the feasibility set, such as cutting plane and extreme point ranking methods \cite{floudas1995quadratic}, can be used to efficiently solve Problem \ref{prob:RSEP-EQ}.

\vspace{-0.1cm}
\subsection{Heuristic Solution} \label{sec:greedy}

Although Problem \ref{prob:RSEP-EQ} has lower complexity than Problem \ref{prob:RSEP-QP}, it is still hard to find a solution as the problem grows in size. Therefore,  we design a polynomial solution to Problem \ref{prob:RSEP-QP} by using a heuristic approach. The key idea is to generate a solution that provides good performance while achieving low computational complexity.
Given Problem \ref{prob:RSEP-QP}  maximizes the number of shared RBs, we can allocate as many linked RBs as possible to those MVNOs that request the highest amount of RBs on multiple interfering BSs. Indeed, MVNOs that request the greatest number of resources on different interfering BSs are also expected to produce a high number of linked RBs.
Accordingly, for each MVNO $m$ we define the \textit{linking index} $l_m$ as 
\begin{equation} \label{eq:l_tilde}
    l_m = \sum_{b\in\bss} \sum_{b'\neq b} \min \{ L_{m,b} , L_{m,b'} \} y_{b,b'}
\end{equation}

The linking index is used to sequentially allocate RBs to those MVNOs with the highest linking index. We refer to this procedure as the \textit{Most Linked First} (MLF) procedure, which is illustrated in Algorithm \ref{Alg:greedy} and works as follows:
 \begin{enumerate}
     \item we generate set $\mvnos^G=\mvnos$ for all $m,k\in\mvnos^G$ s.t. $m<k$, then $l_m\geq l_k$;
     \item we start allocating RBs on all BSs in sequential order to the first MVNO in $\mvnos^G$, \textit{i.e.}, the MVNO whose linking index $l_m$ is the highest among all MVNOs in $\mvnos$. When all RBs are allocated to the considered MVNO, say $m'$, we remove it from $\mvnos^G$ and we set $l_{m'}=0$;
     \item if $\mvnos^G = \emptyset$, we stop. Otherwise, we re-execute Step 2 until all MVNOs are assigned to the required RBs.
 \end{enumerate}

Line 4 requires to compute \eqref{eq:l_tilde} which has complexity $\bigO{MB^2}$, while Line 5 has complexity $\bigO{M\log M}$. The \texttt{while} loop at Line 6 has complexity $\bigO{\nrb B M}$. Thus, the complexity of MLF is $\bigO{C}$, where $C = \max \{MB^2, M\log M,\nrb\cdot  B\cdot  M\}$.

\begin{algorithm}[h]
\begin{algorithmic}[1]
\caption{RSEP-MLF}
\label{Alg:greedy} 
\State \texttt{Input} $\bss; \mvnos; \mathbf{Y}; \mathbf{L}$;
\State \texttt{Output} A MLF RBs allocation $\mathbf{x}^G=(x^G_{m,b,n,t})_{m,b,n,t}$;
\State \texttt{Set} $x^G_{m,b,n,t}=0$ for all $m\in\mvnos,b\in\bss,(n,t)\in\rbs$; 
\State Compute the linking index $\mathbf{l}=(l_m)_{m\in\mvnos}$;
\State $\mvnos^G \leftarrow$ Sort $\mvnos$ by $l_m$ in decreasing order;
\While{$\mvnos^G \neq \emptyset$}
    \For{ each BS $b\in\bss$}
    \State Update $x^G_{m,b,n,t}$ by allocating $L_{\mvnos^G(1),b}$ subsequent RBs to MVNO $m$ on BS $b$;
    \EndFor
    \State $\mvnos^G \leftarrow \mvnos^G \setminus \{\mvnos^G(1)\}$;
\EndWhile
\end{algorithmic}
\end{algorithm}

\vspace{-0.1cm}
\subsection{Speeding-up the execution of RSEP-QP and RSEP-EQ}\label{sec:reduc_complex}

Although Problems RSEP-QP and RSEP-EQ have exponential complexity,  two intuitions help reduce their complexity by leveraging specific structural properties of the RSEP.
\subsubsection{Sparsity} \label{sec:sparse}
Let $\mathbf{x}^{OPT}$ be an optimal solution to either Problem RSEP-QP or RSEP-EQ.
If $L_{m,b}=0$ for a given MVNO $m$ on BS $b$, then $x^{OPT}_{m,b,n,t}=0$ for all $n$ and $t$. 
Furthermore, we notice that the complexity of many optimization problems strongly depends on the number of non-zero entries (\textit{i.e.}, the sparsity) of the $\mathbf{Q}$ matrix, as pointed out in \cite{boyd2004convex}. Thus we leverage the particular structure of our problem to reduce the complexity of the two problems by introducing two transformations. 
Specifically, let $m'$ and $b'$ such that $L_{m',b'}=0$, for both RSEP-QP and RSEP-EQ we generate a reduced matrix $\tilde{\mathbf{Q}}$ where we set $Q_{m',b',n,t}=0$ for all $(n,t)\in\rbs$. For RSEP-QP, it suffices to replace the $\mathbf{Q}$ matrix with $\tilde{\mathbf{Q}}$. Instead, to keep the equivalence between RSEP-QP and RSEP-EQ, the objective function of RSEP-EQ should be reformulated as
\begin{equation}
\frac{1}{2} \mathbf{x}^\top (\tilde{\mathbf{Q}} + 2\lambda \mathbf{\tilde{I}}_V) \mathbf{x} - \lambda    
\end{equation}
\noindent
where $\mathbf{\tilde{I}}_V$ is the identity matrix where we set to zero those entries corresponding to the $2$-tuple $(m',b')$.

Note that the two above transformations generate equivalent problems to RSEP-QP and RSEP-EQ and do not impact the optimality of the computed solutions. 
In fact, Constraint \eqref{constr:B1} requires $\sum_{t=1}^T \sum_{n=1}^{\nrb} x_{m',b',n,t} = 0$ when $L_{m',b'}=0$. Since $x_{m',b',n,t}\in\{0,1\}$, we have that $x_{m',b',n,t}=0$ for all $n$ and $t$ associated to the $2$-tuple $(m',b')$. That is, at the optimal solution, $x_{m',b',n,t}=0$ independently of the value of $q_{m',b',n,t}$.  
\subsubsection{RB Aggregation} \label{sec:aggregation}
Let $K=\mathrm{GCD}(\mathbf{L})$ be the greatest common divisor (GCD) among all of the elements in the $\mathbf{L}$ matrix.  We show that Problems RSEP-QP and RSEP-EQ are equivalent to solve the same problems with a scaled RB grid, when given conditions on $K$, $T$ and $\nrb$ are satisfied. Specifically, if $K>1$ and either the number $\nrb$ of RBs or the number $T$ of time slots are proportional to $K$, the available RBs can be aggregated in groups of $K$ RBs so that each group can be seen as a single aggreagated RB. We refer to such a property as \textit{aggregability} of the RSEP. Definition follows below. 

\begin{definition}[Aggregable RSEP]
The RSEP is said to be \textit{aggregable} if $\nrb\!\pmod{K}\!=\!0$ or $T\!\pmod{K}\!=\!0$, where $K\!=\!\mathrm{GCD}(\mathbf{L})\!>\!1$ and $A\!\!\pmod{B}$ is the $A$ modulo $B$ operator.
\end{definition}

In the first case, we scale the number of RBs as $\tilde{N}_{RB} = \nrb/K$. In the second case, we scale the number of time slots as $\tilde{T} = T/K$. That is, for each BS $b\in\bss$, the set $\rbs_b$ of available RBs at $b$ is replaced with an aggregated version of cardinality $|\tilde{\rbs}_b|=\nrb T/K$ where $K$ RBs are grouped together to create a single RB. 
We refer to this low-dimensional RSEP as the \textit{aggregated RSEP}.

\begin{theorem} \label{th:aggregation}
Let the RSEP be aggregable, it is possible to compute an optimal solution to the RSEP by solving the aggregated RSEP.
\end{theorem}
\begin{IEEEproof}
Let $Z=\nrb T$, $K>1$ be the GCD of $\mathbf{L}$, $P$ be the original RSEP problem and $\tilde{P}$ be the aggregated RSEP with $\tilde{T} = T/K$. The proof for the case where we aggregate with respect to $\tilde{N}_{RB} = \nrb/K$ follows the same steps.
First, note that problem $P$ is shift invariant with respect to the indexing of $n$ and $t$. 
Our statement can easily be proven by simply noting that for any given optimal solution $\mathbf{x}^*$, the solution $\mathbf{x}$ with $x_{m,r,1,t} = x^*_{m,r,\nrb,t}$ and $x_{m,r,\nrb,t} = x^*_{m,r,1,t}$ for all $m$, $r$ and $t$ is clearly still optimal as it produces the same number of linked RBs as $\mathbf{x}^*$. 
In general, we can extend this result to any reshape procedure of the RB set $\rbs$ that maintains the cardinality of $\rbs$ constant and equal to $Z$. 
With this feature at hand, 
we will 
show that we can reduce the cardinality of $\rbs$ by a factor $K$ and still achieve equivalence and optimality.




Let $\mathcal{X} \in \mathbb{R}^{\nrb\tilde{T}\times K} $ and $\mathcal{\tilde{X}} \in \mathbb{R}^{\nrb\tilde{T} \times 1}$ be the feasibility sets of  $P$ and $\tilde{P}$, respectively. 
Also, let $f_{Z}(\mathbf{x}): \mathcal{X}\rightarrow\mathbb{N}$ and $f_{Z/K}(\mathbf{x}): \mathcal{\tilde{X}}\rightarrow\mathbb{N}$ be the objective functions of problem $P$ and $\tilde{P}$, respectively.
The optimal solution to $P$ is denoted as $\mathbf{x}^*\in\mathcal{X}$, while the optimal solution to $\tilde{P}$ is denoted as  $\mathbf{\tilde{x}}^*\in\mathcal{\tilde{X}}$.
Due to the optimality of $\mathbf{x}^*$ and $\mathbf{\tilde{x}}^*$, we have that $f_{Z}(\mathbf{x}^*)\geq f_{Z}(\mathbf{x})$ for all $\mathbf{x}\in\mathcal{X}$, and $f_{Z/K}(\mathbf{\tilde{x}}^*)\geq f_{Z/K}(\mathbf{\tilde{x}})$ for all $\mathbf{\tilde{x}}\in\mathcal{\tilde{X}}$.
Let us define $\mathbf{\tilde{x}}^*_A \in \mathcal{X}$ be the solution to $P$ generated by expanding the aggregated optimal solution  $\mathbf{\tilde{x}}^*$ to $\tilde{P}$. Let $\rbs \in \mathbb{R}^{\nrb\tilde{T}\times K}$, 
$\mathbf{\tilde{x}}^*_A=(\tilde{x}^*_{A_{m,b,\tau,k}})_{m,b,\tau,k}$ is generated by setting $\tilde{x}^*_{A_{m,b,\tau,k}} = \tilde{x}^*_{m,b,\tau}$ for all $k=1,...,K$, $m\in\mvnos$ and $b\in\bss$. Intuitively, we are replicating the matrix $\mathbf{\tilde{x}}^*$ by adding $K-1$ rows whose entries are identical to those in $\mathbf{\tilde{x}}^*$. 

We will now prove that $P$ and $\tilde{P}$ are equivalent by contradiction. Accordingly, we will negate our hypotesis and we will assume that the two problems are not equivalent, i.e., $f_{Z}(\mathbf{x}^*) > f_{Z}(\mathbf{\tilde{x}}^*_A)$.

Let $g(\mathbf{x}): \mathcal{X}\rightarrow\mathbb{N}$ be defined as $g(\mathbf{x})=K^{-1}f(\mathbf{x})$. Intuitively, if we replace the objective function $f(\mathbf{x})$ of $P$ with $g(\mathbf{x})$, we obtain the same problem where each linked RB gives a reward equal to $K^{-1}$ ($f(\mathbf{x})$ instead provides a unitary reward for each linked RB).
By construction of $\mathbf{\tilde{x}}^*_A$, we have $f_{Z/K}(\mathbf{\tilde{x}}^*) = K^{-1} f_{Z}(\mathbf{\tilde{x}}^*_A)=g(\mathbf{\tilde{x}}^*_A)$. 
From the assumption $f_{Z}(\mathbf{x}^*) > f_{Z}(\mathbf{\tilde{x}}^*_A)$, we have that 
\begin{equation} \label{eq:contradiction}
     g(\mathbf{x}^*) \!=\! K^{-1}\! f_{Z}(\mathbf{x}^*)\! > \!K^{-1} \!f_{Z}(\mathbf{\tilde{x}}^*_A)\! =\! g(\mathbf{\tilde{x}}^*_A) \!= \!f_{Z/K}(\mathbf{\tilde{x}}^*)
\end{equation}
\noindent
which states that $g(\mathbf{x}^*) > f_{Z/K}(\mathbf{\tilde{x}}^*)$. 

To show that this last statement is a contradiction to our hypothesis (i.e., $\mathbf{\tilde{x}}^*$ is optimal for $\tilde{P}$), we need to show that there always exist a mapping that transforms any solution in $\mathcal{X}$ to an equivalent solution in $\mathcal{\tilde{X}}$. That is, we need to find a function $h(\mathbf{x}): \mathcal{X} \rightarrow \mathcal{\tilde{X}}$ such that $h(\mathbf{x}) = \mathbf{\hat{x}} \in \mathcal{\tilde{X}}$ that can be transformed into $\mathbf{\hat{x}}_A$ such that $f_{Z/K}(\mathbf{\hat{x}}) = K^{-1} f_{Z}(\mathbf{\hat{x}}_A)$.

In general, such a mapping is not unique, since any optimal solution in $\mathcal{X}$ and $\mathcal{\tilde{X}}$ is shift invariant. However, in Appendix A we present an easy mapping $h(\mathbf{x})=\mathbf{\tilde{x}}$ that, starting from an optimal solution $\mathbf{x}\in\mathcal{X}$, always generates an equivalent optimal solution $\mathbf{\tilde{x}}\in\mathcal{\tilde{X}}$ such that $f_{Z/K}(\mathbf{\tilde{x}}) = K^{-1} f_{Z}(\mathbf{x})$.

The existence of the above mapping implies that $\mathbf{x}^*_R = h(\mathbf{x}^*)$ satisfies $f_{Z/K}(\mathbf{x}^*_R) = K^{-1} f_{Z}(\mathbf{x}^*) = g(\mathbf{x}^*)$, which is clearly a contradiction. In fact, from \eqref{eq:contradiction} we have that $f_{Z/K}(\mathbf{x}^*_R) = g(\mathbf{x}^*) > f_{Z/K}(\mathbf{\tilde{x}}^*)$, which implies the existence of a solution $\mathbf{x}^*_R$ that contradicts the optimality of $\mathbf{\tilde{x}}^*$ over $\mathcal{\tilde{X}}$. It follows that $f_{Z/K}(\mathbf{\tilde{x}}^*) = K^{-1} f_{Z}(\mathbf{x}^*)$ must hold. Hence, any solution $\mathbf{\tilde{x}}^*$ to the aggregated RSEP can be expanded to obtain $\mathbf{\tilde{x}}^*_A$ that is optimal for the original RSEP. This concludes the proof.
\end{IEEEproof}

%% file: numerical_analysis.tex
\vspace{-0.1cm}
\section{Numerical Analysis}\label{sec:numerical}

We now assess the performance of the algorithms proposed in Section \ref{sec:solution}. To this end, we simulate an LTE frequency division duplexing (FDD) system with $\mathrm{1.4~MHz}$ channel bandwidth, which is divided into $72$ subcarriers organized into $N_{RB}=6$ physical resource blocks (PRBs). Each PRB represents the minimum scheduling unit and consists of $12$ subcarriers and $14$ symbols. Time is divided into discrete time slots called \textit{sub-frames}, the duration of each sub-frame equals the duration of one PRB and $N_{SF}=10$ sub-frames constitute a \textit{frame}. Let $N_F\in\mathbb{N}$ be the number of frames within the slicing enforcing window. It follows that $T = N_F\cdot N_{SF}$.

Unless stated otherwise, we assume that both the interference matrix $\mathbf{Y}$ and the slicing profile matrix $\mathbf{L}$ are generated at random at each simulation run. Results were averaged over 1000 independent simulation runs.

\vspace{-0.1cm}
\subsection{Convergence Time Analysis}

\begin{figure}[h]
    \centering
    \includegraphics[width=\columnwidth]{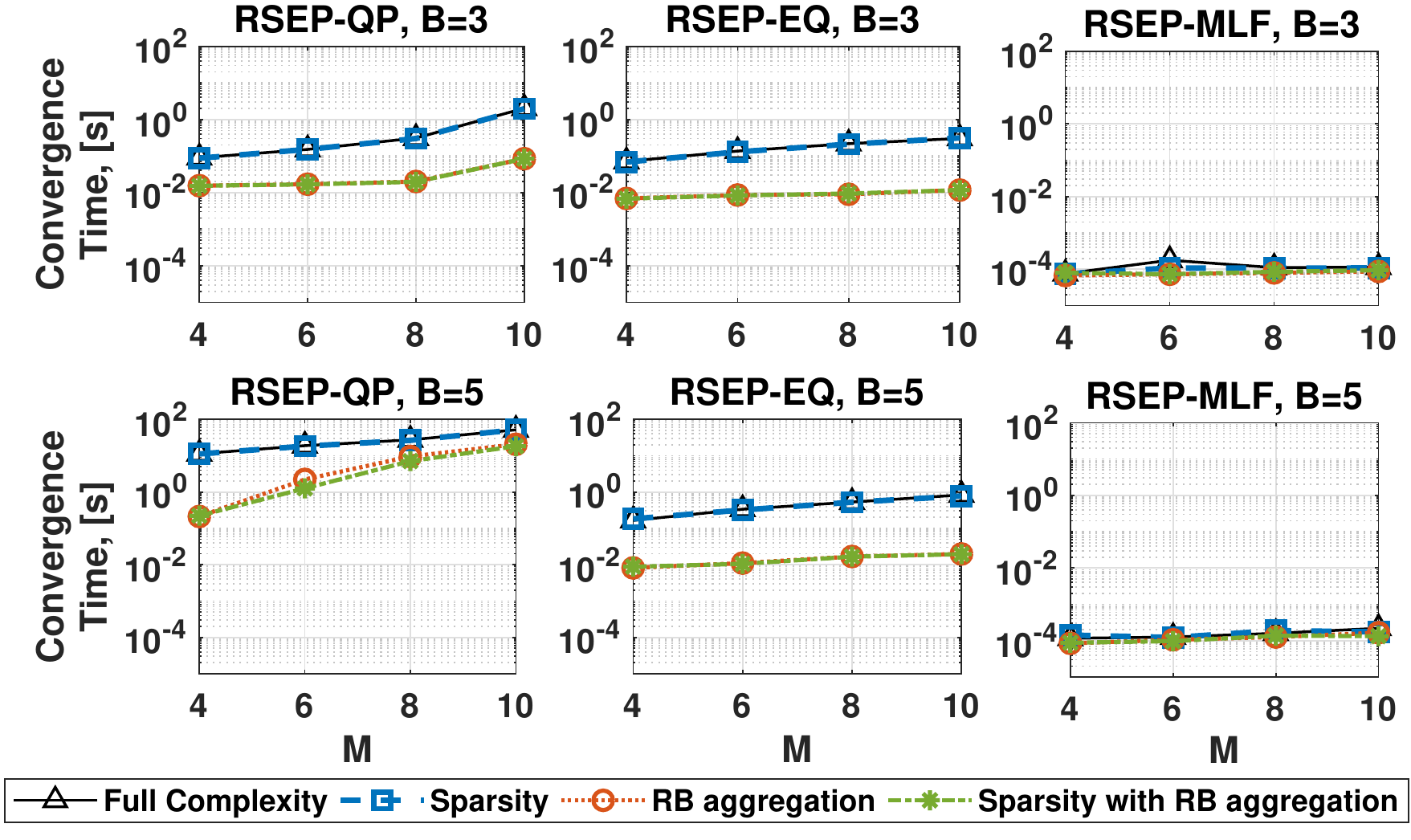}
    \caption{Convergence time (s) of the three proposed solutions as a function of $M$ considering different computational time reduction techniques.}
    \label{fig:convergencetime}
 \end{figure}

\reffig{fig:convergencetime} shows the convergence time of RSEP-QP, RSEP-EQ and RSEP-MLF as a function of the number $M$ of MVNOs when $B=5$ and $N_F = 2$. As expected, the convergence time increases as the number of MVNOs in the network increases. Moreover, the convergence time of RSEP-QP is considerably higher than the one of RSEP-EQ and RSEP-MLF.
\reffig{fig:convergencetime} also shows the impact of the sparsity and RB aggregation mechanisms in Sections \ref{sec:sparse} and \ref{sec:aggregation} on the overall convergence time. As it can be observed, the techniques presented in Section \ref{sec:reduc_complex} can effectively reduce the computation time of all the three problems. Moreover, we show the RB aggregation produces the best performance improvement in terms of convergence time.

\textit{We point out that RSEP-QP requires approximately $100$s to compute an optimal solution when $M=10$ and $B=5$. On the contrary, RSEP-EQ only requires $1$s while RSEP-MLF computes the solution in less then a millisecond.}


Interestingly, \reffig{fig:convergencetime} reveals that the reduction in terms of convergence time brought by sparsity can not be appreciated in small-scale scenarios. For this reason, we have further investigated the impact of sparsity in large-scale networks and the obtained results are presented in \reffig{fig:convergencetime_EQ}. It is shown that sparsity can effectively reduce the computation time by several tens of seconds, and the gain increases as both $M$ and $B$ increase.
 
 \vspace{-0.1cm}
 \subsection{Optimality-gap Analysis}
 
\begin{figure}[h]
\centering
    \begin{minipage}[b]{0.45\columnwidth}
    \includegraphics[width=0.9\columnwidth]{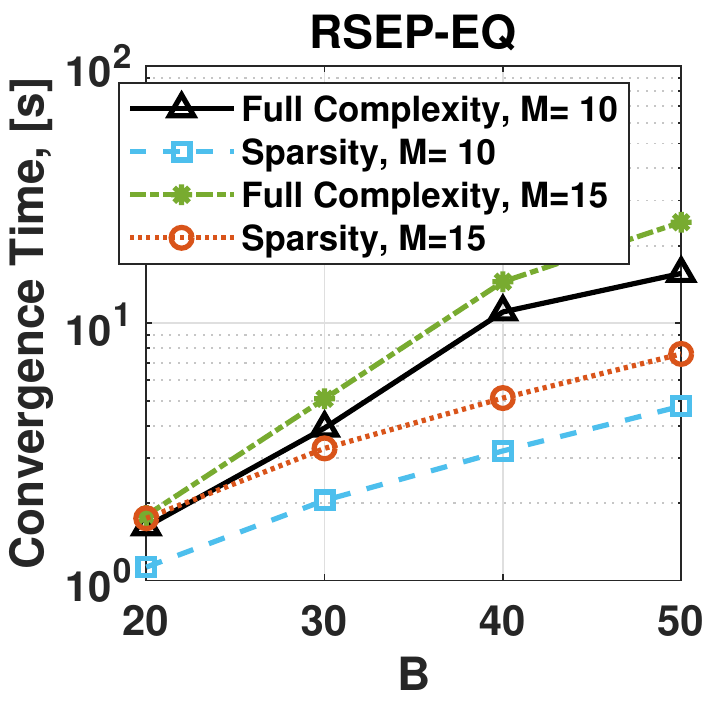}
    \vspace{-0.2cm}
    \caption{\label{fig:convergencetime_EQ} Convergence time (s) of RSEP-EQ as a function of $B$ considering different number $M$ of MVNOs.}
  \end{minipage}
    \hspace{0.03\textwidth}
   \begin{minipage}[b]{0.45\columnwidth}
    \centering
    \includegraphics[width=0.9\columnwidth]{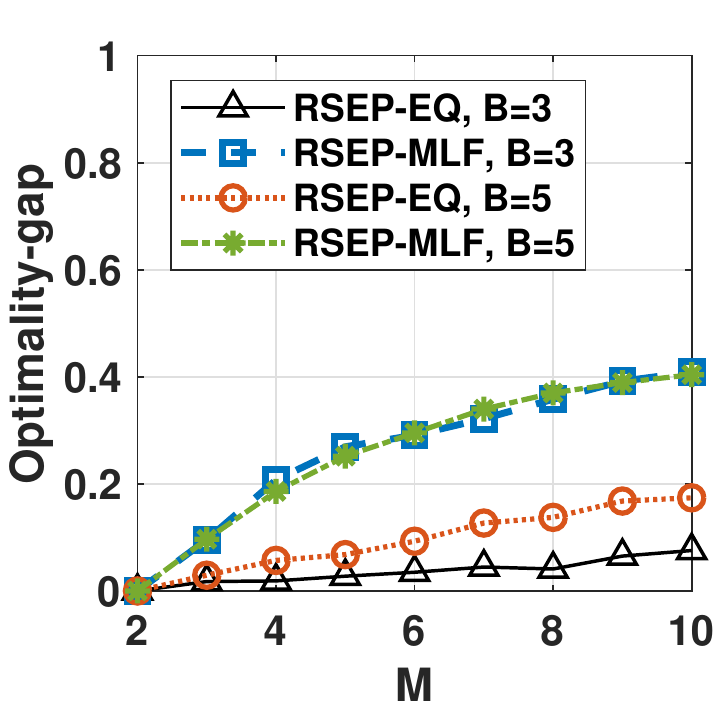}
    \vspace{-0.2cm}
    \caption{\label{fig:optimalitygap} Optimality-gap of RSEP-EQ and RSEP-MLF as a function of $M$ considering different number $B$ of BSs.}
  \end{minipage}
\end{figure}

 Another crucial aspect is the optimality-gap between RSEP-QP and RSEP-EQ/RSEP-MLF. Although Theorem \ref{th:equivalence} shows that (under some conditions) Problem RSEP-EQ is equivalent to Problem RSEP-QP, we can not guarantee that the solution computed by RSEP-EQ is a global optimum. Indeed, the solver might get stuck in one of the local maximizers, thus effectively preventing the computation of an actual global maximizer. Thus, in \reffig{fig:optimalitygap} we investigate the optimality-gap of RSEP-EQ and RSEP-MLF with respect to an optimal solution computed by RSEP-QP. In other words, the closer to zero is the optimality-gap, the closer to optimality the solution is. 
\vspace{-0.1cm}
 
 \reffig{fig:optimalitygap} shows that the optimality-gap increases as the number of MVNOs and BSs in the network increases. Intuitively, this is because the feasibility set increases as $M$ and/or $B$ increase. Given that local maximizers of RSEP-EQ lie on the vertices of the feasibility set, greater values of $M$ and $B$ produce a greater number of local maximizers, thus the probability of getting stuck in a local maximizer increases. 
 Notice that although RSEP-MLF is negligibly affected by the number of BSs $B$, it achieves poor performance if compared to RSEP-EQ. Figs. \ref{fig:optimalitygap} and \ref{fig:convergencetime} also indicate that RSEP-EQ represents an effective solution to the RSEP that is both near-optimal and computed in few seconds. 
 
 \vspace{-0.1cm}
\subsection{Linked RBs Analysis}

\reffig{fig:utility_M} shows the impact of $M$ and $B$ on the total number of linked RBs of the system when $N_F = 10$ and $T = 100$. As expected, RSEP-EQ always performs better than RSEP-MLF in terms of number of linked RBs. 
Moreover, \reffig{fig:utility_M} illustrates that larger values of $B$ produce a greater number of linked RBs. Conversely, the number of linked RBs decrease as the number $M$ of MNVOs increase. This is because, when more MVNOs include the same BS to their slices, it is harder to guarantee that all MVNOs will receive the corresponding amount of RBs joint with a large number of linked RBs.


%% file: experimental_analysis.tex
\vspace{-0.1cm}
\section{Experimental Analysis}\label{sec:experimental}

In this section, we describe our testbed analysis and the results obtained through experimental evaluation. First, we describe in Section \ref{sec:exp_test} our experimental testbed and the scenario considered. Finally, we discuss the experimental results in Section \ref{sec:exp_results}.





\vspace{-0.1cm}
\subsection{Experimental Testbed and Network Scenarios}\label{sec:exp_test}

To evaluate the performance of our algorithms, we have used an orthogonal frequency division multiple access (OFDMA) system as in LTE \cite{astely2009lte}.  In OFDMA, time and frequency are divided into RBs, where each RB has time duration equal to the duration of an OFDMA frame, which spans $K$ subcarriers split into $\Delta T$ slots. Each of these slots are assigned to one MVNO according to the outcome of the slicing enforcement algorithm, who in turn can decide to assign subcarriers to MUs according to its own internal resource allocation and scheduling policy.     

\begin{figure}[h]
\centering
  \begin{minipage}[b]{0.34\columnwidth}
  \centering
  \vspace{-0.2cm}
    \includegraphics[width=\columnwidth]{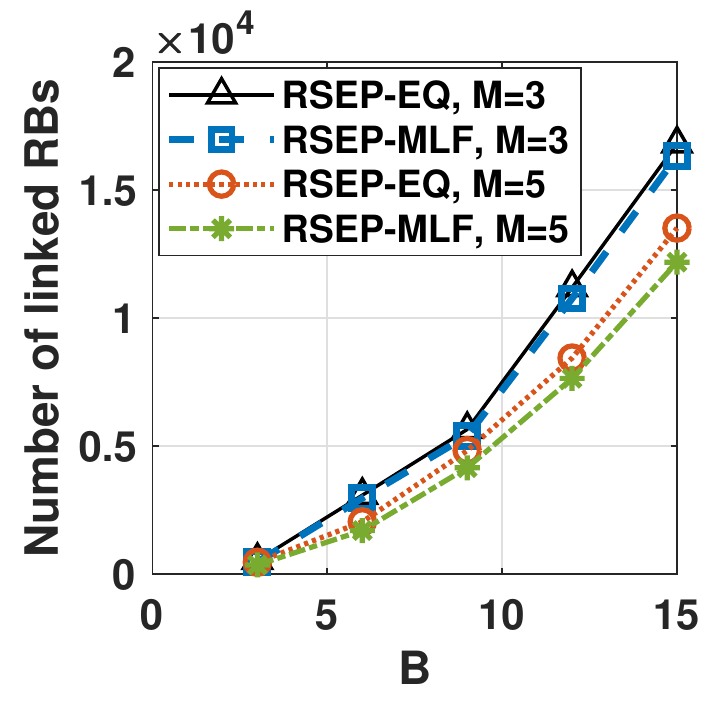}
    \vspace{-0.6cm}
    \caption{\label{fig:utility_M} Total number of linked RBs for RSEP-EQ and RSEP-MLF as a function of $B$ for different values of $M$.}
  \end{minipage}
    \hspace{0.01\textwidth}
   \begin{minipage}[b]{0.62\columnwidth}
    \centering
    \vspace{-0.3cm}
    \includegraphics[width=\columnwidth]{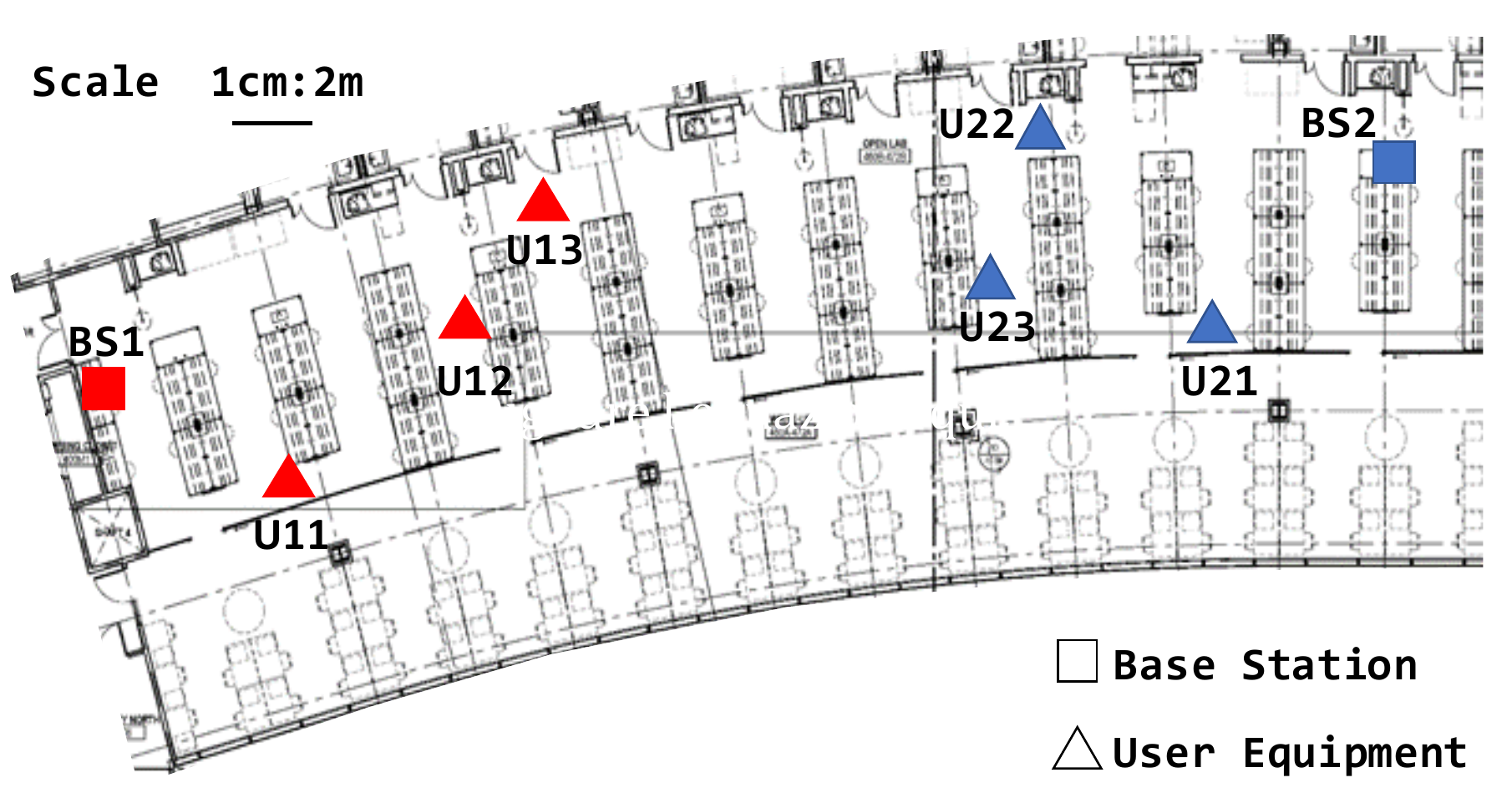}
    \vspace{-0.6cm}
    \caption{\label{fig:exp_testbed} Experimental Testbed.}
  \end{minipage}
  \hspace{0.01\textwidth}
\end{figure}  

 
We set up a testbed of 8 USRP software-defined radio for our experiments.Two USRP X310 act as BSs, 6 USRP N210 are used to implement MUs and all of them run GNU Radio. The USRP radios were deployed as shown in Figure \ref{fig:exp_testbed}, where we show that two groups of three N210s are associated with each BS. The BSs were synchronized in time and phase by using an \textit{Octoclock} clock distributor by Ettus Research.

\subsection{Experimental Results}\label{sec:exp_results}

Our experiments were targeted to evaluate two critical performance parameters:

\begin{enumerate}
    \item the effectiveness of our slicing enforcement algorithms in assigning the spectrum resources (i.e., the RBs) to the MVNOs according to the slicing policy; 
    \item the performance (\textit{i.e.}, total network throughput) of RSEP-QP as opposed to sub-optimal algorithms.
\end{enumerate}

To address point 1), we consider a network with three MVNOs and 6 MUs, where each MU is associated with a different MVNO in each BS (e.g., $U_{11}$ is associated with MVNO 1 and BS1, while $U_{23}$ with MVNO 3 and BS2). We consider the case where the two BSs  enforce the following slicing policies on the three MVNOs, which change every  $\mathrm{T = 300s}$, by using the RSEP-QP algorithm.

\begin{enumerate}
    \item BS1: \{70, 30, 0\}\%;   \{25, 25, 50\}\%; \{30, 40, 30\}\%.  
    \item BS2: \{25, 25, 50\}\%;   \{30, 40, 30\}\%; \{70, 30, 0\}\%.
\end{enumerate}
\begin{figure}[!h]
    \centering
    \includegraphics[width=0.7\columnwidth]{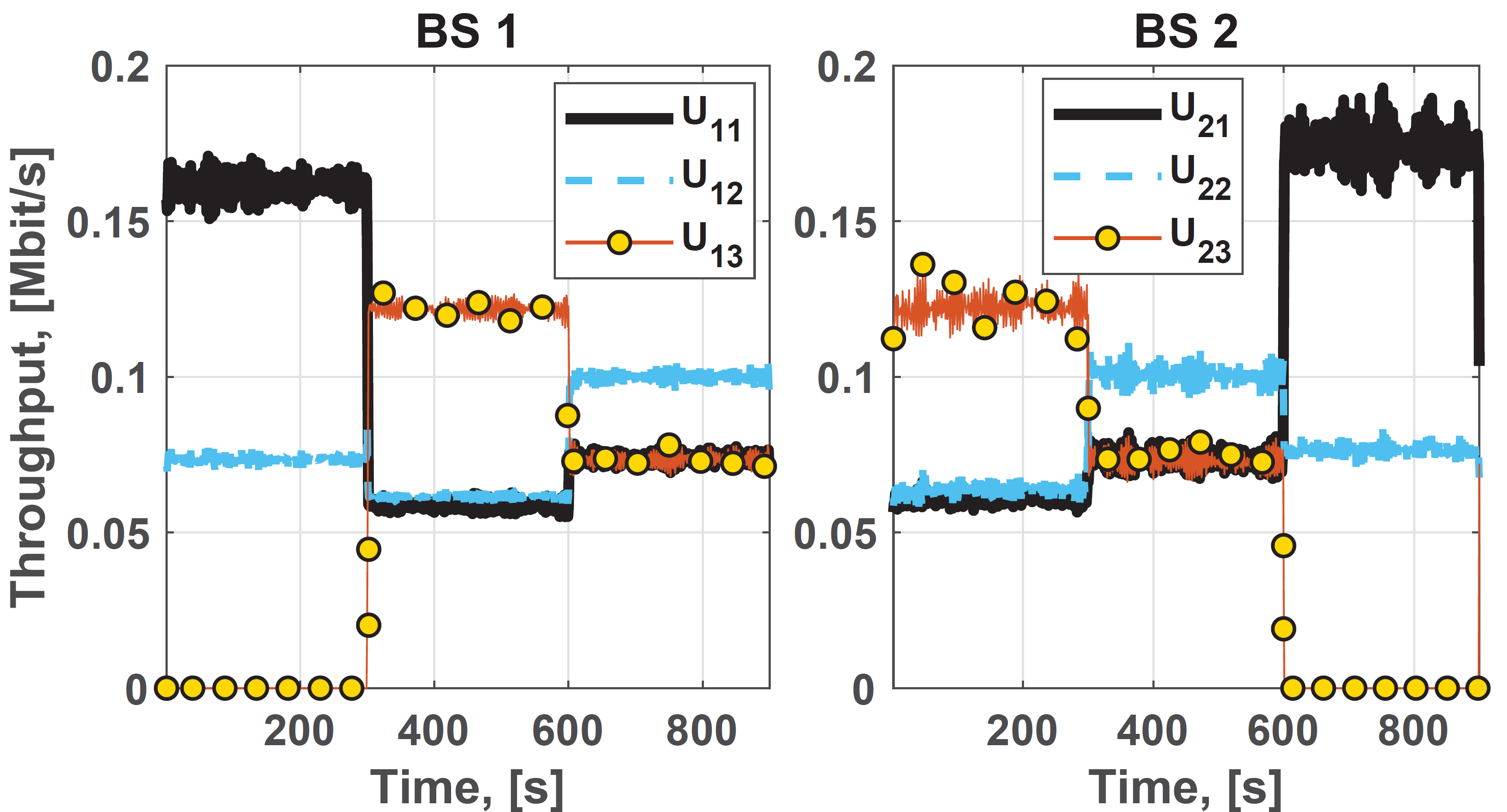}
    \caption{Throughput per MU as a function of time and MUs.}
    \label{fig:first_exp}
 \end{figure}

Figure \ref{fig:first_exp} depicts the throughput experienced by each MU as a function of time at BS1 and BS2, averaged over 10 repetitions. The sharp change in throughput corresponding to different slicing policies in Figure \ref{fig:first_exp} indicates that our RSEP-QP slicing enforcement algorithm indeed assigns to each MVNO (and therefore, each MU) a number of RBs that is coherent to what expressed in the slicing policy.
\vspace{-0.1cm}
\begin{figure}[!h]
    \centering
    \includegraphics[width=0.8\columnwidth]{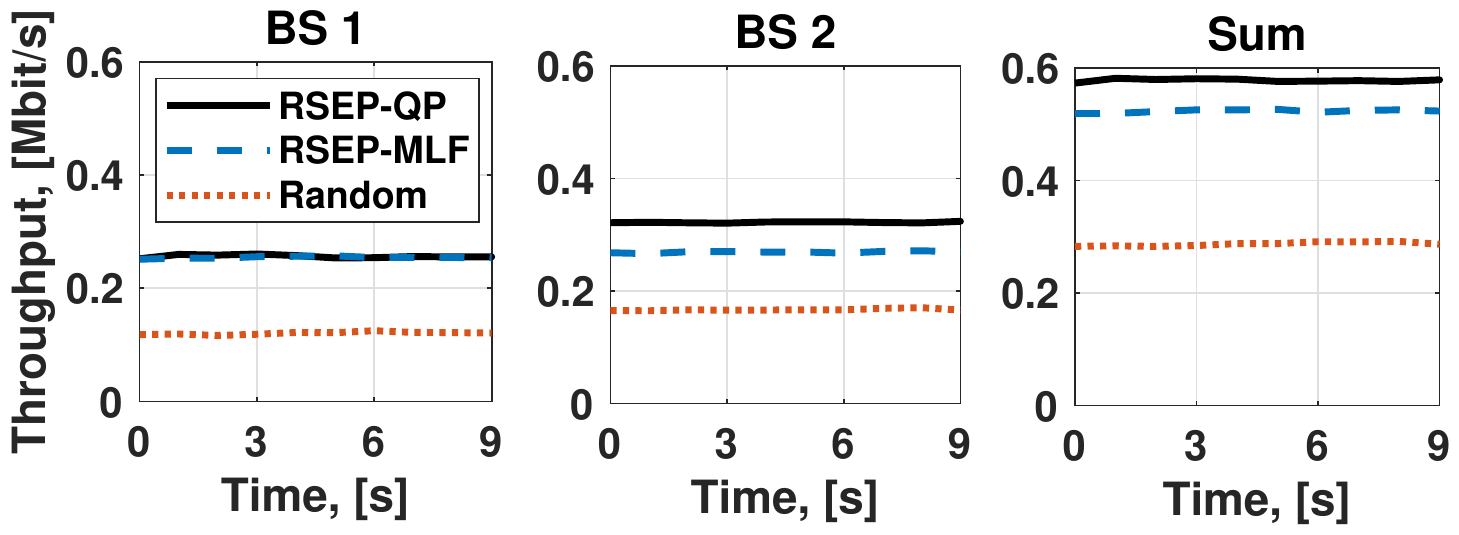}
    \caption{Total network throughput for different slicing enforcement algorithms.}
    \label{fig:sum_thr}
 \end{figure}

In addressing point 2), we consider the same network scenario as in 1). However, to remark the difference between the optimum and the sub-optimal algorithms, we consider 9 MVNOs instead of 3. The MVNOs have been assigned the following slicing policy at the two BSs:

\begin{enumerate}
    \item BS1: \{7, 10, 18, 7, 3, 6, 23, 14, 12\}\%;
    \item BS2: \{5, 3, 16, 13, 10, 5, 30, 7, 11\}\%. 
\end{enumerate}

Furthermore, we consider the following IBSPC policy: when two MVNOs have matching RBs at the two BSs, we compute a power control policy that reduce the transmission power to keep the mutual interference below a fixed threshold. 

Figure \ref{fig:sum_thr} shows the total network throughput as a function of time, for the RSEP-QP, RSEP-MLF and \emph{Random} policies, averaged over 100 repetitions. The results in Figure \ref{fig:sum_thr} indicate that (i) as remarked in the Introduction, the throughput increase by using IBSPC in conjunction to an optimal slicing enforcement algorithm can help increase the total network throughput dramatically with respect to approaches that assign RBs to MVNOs in sub-optimal fashion. Our experiments confirm that RSEP-QP doubles the throughput with respect to the \emph{Random} baseline, and that our RSEP-MLF algorithm approximates well the RSEP-QP algorithm. These results are a consequence of the fact that the RBs shared by the MVNOs are respectively 96, 79, and 13 in the case of RSEP-QP, RSEP-MLF and \textit{Random}, therefore, the opportunity of IBSPC is significantly higher in the former two.


%% file: conclusions.tex
\section{Conclusions}\label{sec:concl} 
\vspace{-0.1cm}
In this paper, we have investigated the challenging and timely problem of radio access network (RAN) slicing enforcement in 5G networks. First, we have formulated the resource slicing enforcement problem (RSEP) and shown its NP-hardness. Then, we have proposed two approximation algorithms that render the problem tractable and scalable as the problem increases in complexity. Finally, we have evaluated the algorithms through extensive simulation and experimental analysis on a real-world testbed composed by 8 software-defined radios ang GNU Radio. Results conclude that our algorithms are scalable and provide near-optimal performance. Moreover, our solutions effectively enforce RAN slicing policies by satisfying MVNOs requirements and by reducing inter-MVNO interference. 

%% file: appendix_shorter.tex
\section{Appendix}
\normalsize
\subsection{Aggregation map from $\mathcal{X}$ to $\mathcal{\tilde{X}}$}\label{appendix:algorithm}
Let us consider the reshaped RB grid $\rbs\in\mathbb{R}^{\nrb T \times 1}$ and let us define the RBAM $\bs{\sigma}=(\bs{\sigma}_b)_{b\in\bss}$ where $\bs{\sigma}_b(\mathbf{x}^*) = (\sigma_{b,\tau})_{\tau\in\rbs}$. Henceforth, $b$ and $\tau$ will represent rows and columns of $\bs{\sigma}$, respectively.
For any given optimal solution $\mathbf{x}\in\mathcal{X}$, we build a map between each RB in $\rbs$ and the MVNO that has been assigned with that RB on BS $b$.
Let $M_{b,\tau}(\mathbf{x}^*)$ be the MVNO that RB $\tau$ has been assigned to, \textit{i.e.}, the MVNO $m$ such that $x_{m,b,\tau} = 1$.
Accordingly, we set $\sigma_{b,\tau} = M_{b,\tau}(\mathbf{x}^*)$.

Let us now introduce some terminology for the sake of simplicity. Two columns $\tau_1$ and $\tau_2$ are said to be \textit{coherently swapped} when all their corresponding entries $\sigma_{b,\tau_1}$ are replaced with those of $\sigma_{b,\tau_2}$ and \textit{vice versa} for all $b\in\bss$.
Two columns are \textit{partially swapped} when only a portion $\hat{\bss}\subset\bss$ of entries is replaced among two columns.
Two entries $\sigma_{b_1,\tau}$ and $\sigma_{b_2,\tau}$ are \textit{linked} if $M_{b_1,\tau}(\mathbf{x}) = M_{b_2,\tau}(\mathbf{x})$ and $y_{b_1,b_2}=1$. Finally, we say that $K$ adjacent entries $\sigma_{b,\tau_1},...,\sigma_{b,\tau_K}$ are \textit{paired} if $M_{b,\tau_1}(\mathbf{x}) = M_{b,\tau_2}(\mathbf{x}) = \cdots = M_{b,\tau_M}(\mathbf{x})$, they are said to be \textit{unpaired} otherwise.

The algorithm works as follow. First,  if any $K$ columns of $\bs{\sigma}$ are identical, i.e., all entries are paired, we remove them from $\bs{\sigma}$ and add one out of those $K$ identical columns to an aggregated RBAM $\bs{\sigma}^A\in\mathbb{R}^{\nrb T/K \times 1}$.  Then, we take the following steps:

\begin{enumerate}
    \item \hspace{-0.1cm}We select the row $b_0$ of $\bs{\sigma}$ with the smallest number of distinct MVNOs and we move it to the lowest row; 
    \item \hspace{-0.1cm}We update $\bs{\sigma}$ by ordering row $b_0$ in MVNO identifier order. This operation (i) creates ordered groups of $K$ entries; and (ii) preserves the optimality of the solution as all columns are coherently swapped;
    \item \hspace{-0.1cm}If all entries in the RBAM $\bs{\sigma}$ have been paired, we stop the algorithm;
    \item \hspace{-0.1cm}If any $K$ columns of $\bs{\sigma}$ are identical, we remove them from $\bs{\sigma}$ and we include one of them to $\bs{\sigma}^A$;
    \item \hspace{-0.1cm}We select the row $b_n$ (among the rows above $b_0$) that shares the highest number of links with $b_0$, and we move it above $b_0$;
    \item \hspace{-0.1cm}If all the entries in $b_n$ are paired, we go to 7); otherwise, we find $K$ unpaired entries and we generate a partial swap of $b_n$ and the upper rows such that i) the number of links remains the same\footnote{Note that the initial unpaired solution is optimal, and any partial swap produces a number of links that is at most as high as that of the initial solution.}; and ii) the $K$ entries are paired. \textit{Since we are forcing the number of links to be the same, any partial swap generated in this step maintains the optimality of the solution. Although the partial swap might change the number of links per tenant, it does not changes the total number of links. Thus, the solution generated by the partial swap and the initial optimal solution are equivalent}; 
    \item \hspace{-0.1cm}We set $b_0 = b_n$ and go to 3).
\end{enumerate}

Upon termination, the algorithm creates the aggregated RBAM $\bs{\sigma}^A$ that is then transformed into a reshaped one $\bs{\sigma}^R$ by replicating its columns exactly $K-1$ times. 
All the entries in the reshaped RBAM $\bs{\sigma}^R$ are paired and the total number of links is equal  to the original optimal RBAM $\bs{\sigma}$. It is easy to note that both RBAMs generate the same number of links, \textit{i.e.}, the aggregation mapping generates an aggregated RBAM that is optimal for the RSEP.